\documentclass[conference]{IEEEtran}
\IEEEoverridecommandlockouts
\usepackage{cite}
\usepackage{amsmath,amssymb,amsfonts}
\usepackage{algorithmic}
\usepackage{graphicx}
\usepackage{textcomp}
\usepackage{xcolor}
\def\BibTeX{{\rm B\kern-.05em{\sc i\kern-.025em b}\kern-.08em
    T\kern-.1667em\lower.7ex\hbox{E}\kern-.125emX}}
\newcommand{\rqanswer}[1]{%
\vspace{0.5em}
\noindent\fbox{%
\begin{minipage}{\dimexpr\columnwidth-2\fboxsep-2\fboxrule\relax}
\textbf{RQ answer.} #1
\end{minipage}}%
\vspace{0.5em}
}
\begin{document}

\title{AI-assisted Script Management for Requirements Elicitation Interviews}



\author{
    \IEEEauthorblockN{Anmol Singhal, Paulo Carvalho, and Travis Breaux}
    \IEEEauthorblockA{Carnegie Mellon University, Pittsburgh, USA\\ 
    \{singhal2, pcarvalh, tdbreaux\}@andrew.cmu.edu}
}

\maketitle

\begin{abstract}
Requirements elicitation interviews require interviewers to balance topic coverage, active listening, and adaptive probing while responding to stakeholders in real time. Although prior work has explored AI support for isolated interviewing tasks, such as script generation and follow-up question generation, little is known about how integrated support affects the interview and what requirements artifacts emerge. Furthermore, script management---which helps the interviewer track topic coverage in real time and decide when to probe further---remains underexplored. This paper presents an AI-assisted elicitation workflow that combines theory-guided script generation grounded in business goals with live support for topic coverage tracking and on-demand follow-up question generation. We evaluate the workflow in a between-subjects quasi-experimental study comparing a \textit{no-training, AI-assisted} condition with a \textit{training, AI-unassisted} condition. Based on a rubric derived from elicitation best practices, the AI-generated scripts score higher than training-only scripts (92.8 vs.\ 74.8 out of 100). AI-assisted interviews cover fewer topics (9.6 vs.\ 14.5), cover more scripted questions (86\% vs.\ 69\%), ask more follow-ups per topic (3.43 vs.\ 1.15), and produce more refined goal models (lowest-level goal fraction 0.653 vs.\ 0.598). Participants find script management useful, rating topic tracking as the most useful workflow feature (86\% agreement). Collectively, these results show that the AI-assisted condition is associated with a different interview trajectory and different elicited requirements than a training-only condition, positioning AI-assisted workflows as elicitation scaffolds for future studies. 


\end{abstract}


\section{Introduction}

\begin{figure}[t]
\centering
\includegraphics[width=\columnwidth]{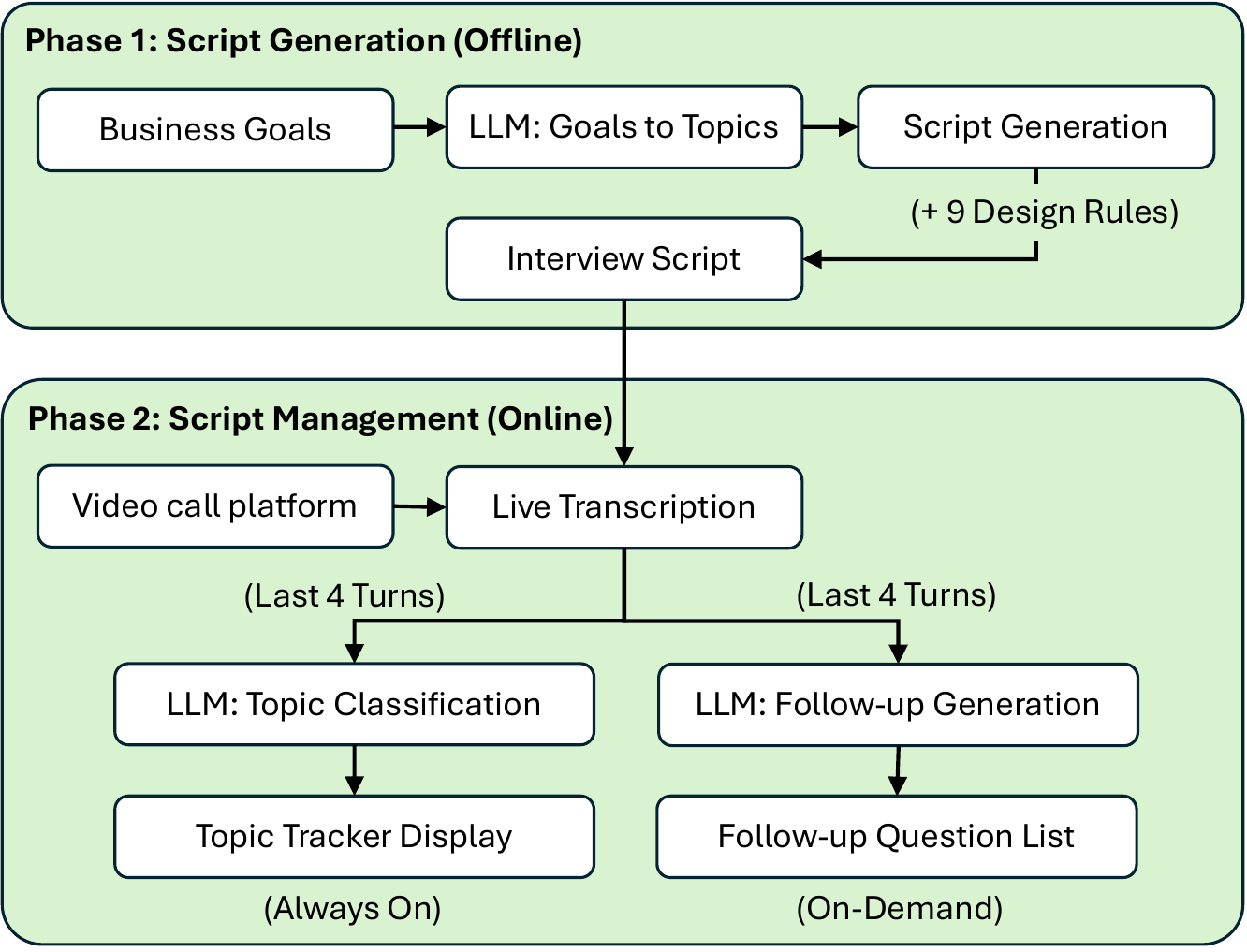}
\caption{The AI-assisted elicitation workflow: offline script generation (Phase~1) followed by online script management (Phase~2).}
\label{fig:workflow}
\vspace{-5mm}
\end{figure}

Understanding stakeholder needs is a persistent challenge in software engineering (SE). When requirements are unclear, incomplete, or misunderstood, teams often face defects, project delays, and expensive rework~\cite{sommerville1999benefitsRE}. Requirements elicitation interviews address this problem by enabling UI/UX designers, product managers, and software developers to collect, clarify, and negotiate requirements with stakeholders~\cite{zowghiRequirementsElicitationSurvey2005, davisEffectivenessRequirementsElicitation2006}. 



Conducting an elicitation interview is challenging because interviewers must simultaneously listen, interpret domain-specific information, decide when to probe, and maintain a productive conversation~\cite{palomaresStateofpracticeRequirements2021}. To support this process, interviewers typically prepare an interview script consisting of questions they believe are relevant to the interview goal. During the interview, however, they must balance following this script with adapting it in response to new information from stakeholders. This requires continuously monitoring topic coverage while deciding whether to ask follow-up questions or shift the conversation, creating substantial cognitive load~\cite{hanwayEffectsCognitiveLoad2020}. Under these conditions, prior studies show that interviewers often ask generic or leading questions, miss tacit assumptions, fail to resolve ambiguities, or struggle to cover relevant topics~\cite{donatiCommonMistakesStudent2017,banoLearningMistakesEmpirical2018}. As a result, interviews may appear successful while important requirements remain undiscovered.


Therefore, interviewers need support not only before the interview, when scripts are prepared, but also during the interview, when they must balance topic coverage, active listening, and adaptive probing under time pressure. Requirements engineering (RE) training introduces interviewers to best practices for preparing and conducting interviews, but provides little assistance once the interview is underway~\cite{ferrariLearningRequirementsElicitation2019, ferrariSaPeerReverseSaPeer2020, mohedasRecommendedInterviewingPractices2022}. Research based on large language models (LLMs) has yielded techniques for generating interview scripts~\cite{parkerGuidelinesIntegrationLarge2023, gorerGPTPoweredElicitationInterview2024}, generating follow-up questions~\cite{shenRequirementsElicitationFollowup2025}, eliciting requirements from LLMs~\cite{ronankiInvestigatingChatGPTPotential2023}, and extracting requirements from interview transcripts~\cite{sharfuddinGenerativeGoalModeling2025}. However, this prior work has largely been conducted independently of an interview or on post-interview transcripts without measuring the impact on interviews. Therefore, we respond to two unexplored research gaps: continuous assistance that (1) tracks which scripted topics have been covered; and (2) supports the interviewer in deciding when to probe a stakeholder response or advance to their next question. Second, we lack outcome-level evidence on how elicitation support techniques change interview trajectories and what requirements are produced. 


We present an AI-assisted elicitation workflow (see Figure~\ref{fig:workflow} and Section~\ref{sec:assistant}) designed to address these gaps. The workflow has two phases: an offline \emph{script generation} phase, wherein an interview script is produced, grounded in business context instead of the interviewer's prior assumptions; and an online \emph{script management} phase that tracks topic coverage and generates on-demand follow-up questions during interviews.

We evaluate the workflow by conducting a between-subjects quasi-experiment comparing two settings (see Section~\ref{sec:evaluation}): an AI-assisted condition without training, in which interviewers use the workflow but do not receive elicitation interview training; and a training-only condition without workflow assistance, in which interviewers conduct interviews by monitoring their own manually prepared scripts. This experimental design does not intend to isolate AI assistance from training as a causal factor, instead it compares how untrained engineers perform with an AI-assisted workflow to engineers who receive conventional elicitation interview training in a classroom setting.

Our evaluation measures three facets of elicitation: (i) script quality using a rubric derived from prior elicitation literature, (ii) interaction quality using speaker turn-based measures of script coverage and follow-up questioning, and (iii) the scale of elicited requirements refinement, measured using goal models extracted from interview transcripts. Goal models are a requirements artifact that can be used to refine high-level goals into low-level system operations~\cite{anton1996gore, lamsweerde2001gore}. 

Our findings (see Section~\ref{sec:results}) indicate that AI-generated scripts score substantially higher than human-authored scripts on quality. AI-assisted interviews achieve more script coverage, focus on fewer topics, and yield goal models with a higher proportion of low-level requirements. Together, the findings present the AI-assistance workflow as increasing interaction focus and volume with positive usefulness perceived by interviewers. The results include implications for practitioners and researchers reported in Section~\ref{sec:discussion}. Threats to validity appear in Section~\ref{sec:threats-validity} with our conclusion in Section~\ref{sec:conclusion}. 


In summary, this paper presents the following contributions:

\begin{itemize}
\item An AI-assisted elicitation workflow that combines business-goal-grounded, theory-guided script generation with online script management for human-led interviews.
\item An outcome-level evaluation spanning script quality, interaction quality, and elicited requirement refinement.
\item A quasi-experimental study showing that a no-training, AI-assisted condition significantly differs from a training-only condition in script quality, interaction volume and focus, and goal-model depth.
\end{itemize}







\section{Related Work}

This paper builds on established foundations in requirements elicitation and interview automation.


\subsection{Interviews for Requirements Elicitation}

Interviews are widely used in requirements elicitation~\cite{zowghiRequirementsElicitationSurvey2005, davisEffectivenessRequirementsElicitation2006, palomaresStateofpracticeRequirements2021}. However, interview quality depends on the interviewer's ability to manage the conversation in real time~\cite{hadarRoleDomainKnowledge2014, banoLearningMistakesEmpirical2018}. Prior work has categorized different failure modes for elicitation interviews, including interviewer mistakes~\cite{donatiCommonMistakesStudent2017, banoLearningMistakesEmpirical2018, spoletiniInterviewReviewEmpirical2018}, ambiguity in stakeholder responses~\cite{ferrariAmbiguityTacitKnowledge2016}, cognitive biases~\cite{zalewskiCognitiveBiases2020, hadarRoleDomainKnowledge2014} and human aspects such as personality, motivation, and demographics~\cite{hidellaarachchiEffectsHumanAspects2022}. Best-practice has been distilled rubrics~\cite{lendingRubricEvaluateEnhance2022, mohedasRecommendedInterviewingPractices2022}, question type taxonomies~\cite{zarembaTypologyQuestions2021}, and procedural prompts to prepare interviewers~\cite{pittsImprovingRequirementsElicitation2007, wahbehSociotechnicalProcessQuestionnaire2020, burnayWhatStakeholdersWill2014}. Training-focused research uses role-playing~\cite{ferrariSaPeerReverseSaPeer2020, zowghiTeachingRequirementsEngineering2003}, peer review~\cite{ferrariLearningRequirementsElicitation2019}, self-assessment~\cite{ daunSurveyInstructionalApproaches2021}, and serious games~\cite{ibrahimDesignDevelopmentSerious2019, liuBARADynamicStatebased2023} as pedagogical methods. While these efforts establish a strong foundation, they primarily cover offline training prior to conducting the interview or as a post-interview assessment. 

\subsection{Automated and LLM-Based Techniques for RE Interviews}

Automation spans the interview lifecycle. \emph{Pre-interview} work focuses on the use of LLMs for script~\cite{ gorerGPTPoweredElicitationInterview2024} and protocol generation~\cite{parkerGuidelinesIntegrationLarge2023} and evaluating script quality~\cite{ElMaarawiTefur2026Debugging}. \emph{In-interview} support includes follow-up question generation~\cite{shenRequirementsElicitationFollowup2025, zhangHarnessingPowerAI2025} and AI-led interviewing~\cite{kornLLMREIAutomatingRequirements2025}. \emph{Post-interview} work covers requirements extraction from interview transcripts, including use case models~\cite{bajajMUCEMultilingualUsecase2022} and goal models~\cite{chenUseGPT4Creating2023, sharfuddinGenerativeGoalModeling2025}.

\subsection{Gaps in Prior Work}

We identify three gaps in prior work. To our knowledge, no prior work has examined script management during elicitation interviews. Research to support interviews examines the generation of follow-up questions~\cite{shenRequirementsElicitationFollowup2025} or replacing humans with AI interviewers~\cite{kornLLMREIAutomatingRequirements2025}. Human interviewers need assistance with maintaining awareness of which planned topics have been covered, deciding when a topic has been sufficiently explored, and choosing when to probe deeper within a topic. Second, while recent work has explored AI-generated interview scripts, these scripts are derived from prompts, requirements documents, or stakeholder personas. Prior work has yet to ground interview topics in business goals, which drive product development. Software engineers focused on feature development may overlook the criticality of business goals in establishing their company's market competitiveness~\cite{lavazza2013businessgoals}. By grounding scripts in business goals, interview support can promote alignment between elicited requirements and a company's priorities. Third, evaluations of existing AI support for elicitation interviews~\cite{ElMaarawiTefur2026Debugging, shenRequirementsElicitationFollowup2025} do not focus on outcome-level effects, such as the participant interaction quality and the scale of refined requirements produced.

\section{AI-Assisted Elicitation Workflow}
\label{sec:assistant}

We propose an AI-assisted elicitation workflow that supports the interviewer in two distinct phases: (i) an offline \emph{script generation} phase that produces an interview script from business goals; and (ii) an online \emph{script management} phase that runs alongside a live interview conducted via Zoom, tracking which scripted topics have been addressed and proposing follow-up questions that interviewers may view on demand. 


\subsection{Script Generation}
\label{sec:script-gen}

Best practice recommends aligning interview questions with the strategic objectives of the organization that owns the product or service being developed~\cite{lavazza2013businessgoals}. Grounding the topics in external evidence helps mitigate the interviewer's prior knowledge bias, in which interviewers tend to probe only what they already understand~\cite{hadarRoleDomainKnowledge2014, zalewskiCognitiveBiases2020}.

We generate interview scripts using \emph{business goals}---specific strategic objectives that a company seeks to achieve through a product or service. To study this phase, we mine business goals from the public disclosures (US Securities and Exchange 10-K filings) of US companies selling products and services in the same application category that interviewers will cover in their interviews. For example, interviewers who will ask stakeholders about video streaming application features will use scripts generated from business goals mined from the public disclosures of Netflix, Apple TV, and YouTube. The business goal mining method is described in Section~\ref{sec:study-procedure}.

An LLM (\texttt{gpt-5-mini}) translates a given set of business goals into one or more candidate \emph{interview topics} (short noun phrases) describing a feature area or user-facing concern through which a user can achieve or obstruct the underlying goal. For example, the topic ``Personalized home recommendations'' derives from the business goal ``Maximize engagement on personalized dashboard'' for a Video Streaming application. 

Not all business goals translate into topics that are relevant to end users. Generating interview questions for every business goal risks introducing questions that the stakeholder cannot meaningfully answer. Therefore, candidate topics are filtered to maximize the interviewer's \textit{familiarity} with those topics retained for the script. In our study, we operationalize interviewer familiarity using a pre-interview survey (see Section~\ref{sec:study-procedure}). The most familiar topics serve as the input to the script generation prompt.


Next, an LLM (\texttt{gpt-5-mini}) is prompted to generate an interview script. The prompt context includes the interview goal verbatim (see Step 3, Section~\ref{sec:study-procedure}), the generated and filtered interview topics, and script-design rules synthesized from the elicitation literature~\cite{banoLearningMistakesEmpirical2018, mohedasRecommendedInterviewingPractices2022, gorerGeneratingRequirementsElicitation2023}. These rules include aligning questions with the interview goal, including an introduction and a summary at the end, asking context-free questions first, ensuring coherent question ordering, asking open-ended questions, asking one question per turn, not asking long or vague questions, and no technical jargon.

\subsection{Script Management}
\label{sec:script-mgmt}

The script-management phase runs alongside the interview and provides two parallel services: \emph{topic coverage tracking} and \emph{on-demand follow-up question generation}. 


After the interviewer initiates a Zoom call for conducting the interview, an automated Zoom bot joins the call as a participant, receives and forwards the meeting audio to a real-time transcription service. The transcriber emits finalized speech turns between the interviewer and the stakeholder to a transcript processor, which tracks topic coverage. 

\textit{Topic coverage tracking}: Within the processor, incoming turns are aggregated into a sliding processing window. Upon a predetermined window size (four speaker turns), a background process submits the window---together with the interview's scripted topic list---to \texttt{gpt-5-mini} via a topic-classification prompt. The prompt instruction is to mark a topic as \emph{addressed} only when it is explicitly mentioned or substantively discussed in the excerpt, and to ignore brief or merely metaphorical references. In addition, every marked topic should include a transcript excerpt as evidence for coverage. The LLM hyper parameters \texttt{reasoning=low} and \texttt{verbosity=low} were selected to keep per-window latency compatible with the conversational pace of a 20--30 minute interview. Each scripted topic is presented in the interviewer's user interface (UI) with a marking to indicate if the topic coverage was detected.

\textit{On-demand follow-up question generation}: A separate process generates follow-up question suggestions. When the interviewer presses a UI button to show follow-up questions, the four most recent turns are passed as context to \texttt{gpt-5-mini} via a second prompt that produces three suggested follow-up questions. The prompt enforces constraints that address common interviewer mistakes~\cite{donatiCommonMistakesStudent2017, banoLearningMistakesEmpirical2018, shenRequirementsElicitationFollowup2025}: questions must (i) clarify ambiguous or under-specified statements; (ii) probe for concrete detail without introducing topics outside the current script scope; (iii) be open-ended, non-leading, simple, and under 20 words; and (iv) avoid soliciting technical implementation details, focusing instead on user needs and experiences. We adopt the prior finding that four turns were sufficient context for 98\% of human-spoken follow-up questions in unassisted interviews~\cite{shenRequirementsElicitationFollowup2025}. Lastly, we chose to present topic tracking and question generation separately in the UI to avoid presenting unnecessary suggestions that could overload the interviewer's conversational flow and active listening~\cite{liuInterviewAIssistantDesigning2025, hanwayEffectsCognitiveLoad2020}.

\section{Research Design}
\label{sec:study-design}

We conduct a between-subjects quasi-experiment to evaluate whether an untrained interviewer using the AI-assisted workflow produces comparable or different elicitation outcomes compared to a trained interviewer who manually prepares and monitors their own interview script. This study does not isolate the independent causal effects of AI assistance versus training. Instead, it compares two practically relevant settings: historically, business analysts have been trained to prepare and manage elicitation interviews manually, whereas emerging AI-assisted SE workflows increasingly enable engineers with limited interviewing expertise to interact directly with stakeholders while relying on AI support throughout their work~\cite{storey2025software,zachariasDevelopersDilemmaOpportunities2026}. Although delegating tasks to AI can substantially improve productivity, prior work has also shown that it can reduce the quality of software artifacts when outputs are accepted without appropriate human oversight~\cite{storey2025software}. The design is quasi-experimental rather than fully randomized because the two groups are drawn from distinct but comparable populations of graduate students with at least one year of industrial SE experience.

\subsection{Research Questions}

We investigate three research questions (RQs):
\begin{itemize}
\item \textbf{RQ1. } To what extent do interview scripts in AI-assisted and training-only conditions conform to script design best practices?
\item \textbf{RQ2. } How does the interaction volume, or interviewer turns spent eliciting information, and interaction focus, or the coverage of follow-up questions distributed across topics and questions, differ between AI-assisted and training-only conditions?
\item \textbf{RQ3. } How does the breadth of requirements and refinement depth in elicited requirements differ between AI-assisted and training-only conditions?

\end{itemize}

We present metrics for operationalizing RQs in Section~\ref{sec:evaluation}.



\subsection{Participant Recruitment}
\label{sec:participants}

For the AI-assisted and training-only conditions, we recruit interviewer--stakeholder dyads. Each dyad is assigned the same task: participating in a role-play elicitation interview about an application in an assigned category. For each dyad, the category is chosen to maximize the stakeholder's familiarity with the category. The interviewer's goal is to ask the stakeholder questions about their behavioral patterns and use cases for applications in the category, and what changes the stakeholder would like to see implemented to improve such applications. The stakeholder's goal is to respond to interviewer's questions to the best of their ability. 

The inclusion criteria for recruiting interviewers under both conditions require: (i) at least one year of industry experience in a technical role; and (ii) prior experience interacting with end users or stakeholders of a software product. Participants in both conditions are drawn from graduate-student populations at a US university, but are recruited through different channels:
\begin{itemize}
    \item \textbf{AI-assisted} interviewers are graduate students in computer science and product management programs who are recruited through brochures, mailing lists, and Slack channels and screened against the same inclusion criteria. These participants did \emph{not} receive in-class elicitation training prior to conducting their interviews.
    \item \textbf{Training-only} interviewers are graduate students enrolled in a graduate requirements engineering course at the university. Prior to their interviews, they received a one-hour in-class instruction on elicitation interviews following Ferrari et al.'s role-play training methodology complemented by watching two videos on tips for elicitation interviews\footnote{https://www.youtube.com/watch?v=kjaOuOYsdUM} and avoiding common mistakes\footnote{https://www.youtube.com/watch?v=p3z8eozPzeA}~\cite{ferrariLearningRequirementsElicitation2019, ferrariSaPeerReverseSaPeer2020}. 
    
\end{itemize}

Each interviewer is paired with one \emph{stakeholder} who serves as the interviewee. The stakeholders are experienced end-users of the assigned application category, i.e., they have used the application \textit{at least once a month}. For AI-assisted dyads, stakeholders are recruited from the same graduate student pool through the same flyer, email, and Slack channels used to recruit AI-assisted interviewers. For training-only dyads, the stakeholders are classmates of the same graduate course who had been assigned the stakeholder role for that interview. Within each condition, interviewers are randomly assigned to stakeholders, and no participant serves as both interviewer and stakeholder. Each interview covers one of 19 consumer-facing app categories, including job search, payments and money transfers, programming/IDE platforms, travel booking, etc.

After two pilot sessions to test the study protocol, which were excluded from the study data, the study enrolled 48 interviewer--stakeholder dyads: 24 assigned the AI-assisted condition and the other 24 assigned the training-only condition. Five dyads (3 AI-assisted, 2 training-only) were excluded because they did not complete all required study steps, which yielded a final sample of 21 AI-assisted, 22 training-only.


\subsection{Procedure}
\label{sec:study-procedure}
The study incorporates a mixed-methods design and combines: (i) a pre-interview survey for the interviewers of both conditions, (ii) audio-recorded virtual interviews between an interviewer and a stakeholder over Zoom, and (iii) a post-interview survey for interviewers in the AI-assisted condition. The study protocol was approved by the University's Institutional Review Board (IRB). Together, the procedure includes a business goal data set simulation step, followed by four additional steps per dyad.


\textbf{Step 0: Simulated business goals.}
The workflow's script generation phase requires business goals (see Section~\ref{sec:script-gen}). To simulate business goals in the experiment, we identified a set of two or more publicly listed US companies that operate a product in each app category. Next, we obtained their most recent annual US Securities and Exchange (SEC) 10-K filings.~\footnote{The 10-K filing is an annual report that provides a detailed overview of a public company's business operations, financial health, and risk factors.} For each filing, we extract text from relevant sections (\emph{Business}, \emph{Risk Factors}, \emph{Management's Discussion and Analysis}) and prompt \texttt{gpt-5-mini} to extract a list of candidate business goals for that company's products. Next, we aggregate the generated goal lists for each category using agglomerative clustering over sentence-embeddings (computed using \texttt{paraphrase-MiniLM-L6-v2}) of individual goal statements, yielding eight consolidated business goals per category on average. The final goal list is used in script generation to produce the interview topics for the app category.

\textbf{Step 1: Consent and assignment.}
We distribute a consent form collecting age and language eligibility, audio-recording consent, and two demographic items: years of industry technical experience (4-point ordinal scale: \emph{No experience}, \emph{less than a year}, \emph{1--3 years}, \emph{$> 3$ years}); and frequency of stakeholder interaction in relation to their prior software development experience (5-point ordinal scale from \emph{Never} to \emph{Very Frequently}). Participants complete the consent form prior to assignment to an application category and pairing with a stakeholder by the study team.

\textbf{Step 2: Script preparation.}
The interviewers in both conditions complete a pre-interview survey. The survey asks each interviewer to rate \emph{familiarity} with the business goal-derived topics in their assigned application category from a scale from one (not familiar) to five (extremely familiar). We record these responses so that higher familiarity values indicate greater experience with current application support. Topics marked ``not familiar'' (rating of one) are removed and the remaining topics are ranked according to the familiarity rating. We finally select 7-8 topics with the highest ranking for script generation. In the AI-assisted condition, the selected topics are fed into the script generation phase of the workflow (see Section~\ref{sec:script-gen} for generating the interview script).

While interviewers in the training-only dyads complete the survey, they do not receive AI-generated scripts. Their interviewers prepare scripts manually using in-class training, which covers the script-design practices later scored in RQ1, such as building rapport, opening with context-free questions~\cite{Gause1989Exploring}, including open-ended questions~\cite{potts1994inquiry} and procedural prompts~\cite{pittsImprovingRequirementsElicitation2007}, summarizing requirements at the end, and avoiding common mistake types~\cite{banoLearningMistakesEmpirical2018}. 


\textbf{Step 3: Interview.}
Each interview is conducted over Zoom and audio-recorded. Both groups receive the same interview goal: \emph{learn how a user of a web or mobile application in the assigned category uses one or more features to achieve a goal, and how the application could be extended to improve performance or offer new features.} Interviewers are instructed to target a session length of approximately 20--30 minutes.



\begin{itemize}
    \item \textbf{AI-assisted} interviewers view the script in the workflow's web interface and conduct the interview while the workflow's script management phase runs alongside the Zoom session. Prior to the interview, these interviewers watch a 10-minute recorded walkthrough of the web interface prior to their session, but receive no instruction on elicitation interviewing (see Section~\ref{sec:assistant}).
    \item \textbf{training-only} interviewers conduct the interview using only their manually prepared script. They receive no further assistance during the interview.
\end{itemize}

\textbf{Step 4: Post-interview.} AI-assisted interviewers complete a post-survey that consists of Likert questions to rate the perceived utility of the workflow's capabilities. Stakeholders do not complete a post-interview survey.


The audio recordings from the interviews are transcribed using Zoom and then post-processed by an LLM (\texttt{gpt-5-mini}) to correct for any grammatical errors. We anonymize and manually convert each generated artifact, irrespective of the study condition, into a consistent format to blind evaluators to the condition to which the interviewer is subject. Participants are compensated after completing all four phases of the study (see replication package for details).

\section{Evaluation}
\label{sec:evaluation}

We now elaborate on the metrics used to answer each RQ, including the hypotheses and the statistical tests. 

\subsection{RQ1: Measuring Script Quality}
\label{sec:rq1-metrics}

Research question RQ1 asks whether script-generation produces scripts of equal or different quality than human-authored scripts, wherein quality is defined by conformance to established script design best practices. We answer RQ1 using an independently verified script evaluation rubric derived from prior elicitation literature. The rubric consists of nine rules that include a one-sentence description, examples, a coding guideline, and scoring technique. The rules were applied to each question and topic block in every script: where a rule was applicable, the question or topic block was coded according to the rule's coding guideline; where a rule was non-applicable, a default N/A code was applied. For each rule $r \in \{1,\dots,9\}$, let $I_r$ be the set of items (questions or topic blocks) to which the rule applies, $\ell_i \in \mathcal{L}_r$ the label assigned to item $i$ from the rule's label space $\mathcal{L}_r$, and $C_r \subseteq \mathcal{L}_r$ the set of compliant labels. The rule score $s_r$ is the probability of compliant labels given the distribution of all labels, and the total score is the unweighted mean of the nine rule scores scaled to $[0,100]$:
\begin{equation}
\label{eq:rubric-score}
s_r = \frac{1}{|I_r|}\sum_{i \in I_r}\mathbf{1}\!\left[\ell_i \in C_r\right],
\qquad
\mathrm{Score} = \frac{100}{9}\sum_{r=1}^{9} s_r .
\end{equation}


The first author designed the rubric based on rubric-design ~\cite{Jonsson2007Rubrics, lendingRubricEvaluateEnhance2022} and script-design best practices~\cite{gorerGeneratingRequirementsElicitation2023, banoLearningMistakesEmpirical2018}. The second and third authors, who have 5+ years of experience in rubric design, validated the rubric by applying it to three scripts excluded from the study data, resolving disagreements in their labels, and editing the rubric to clarify instructions.

Following rubric validation, scripts are scored using a \textit{script-quality judge} based on LLM-as-a-judge~\cite{zheng2023llmjudge}. For each script, we prompt \texttt{gpt-5.4-mini} to assign a code to every question and topic block for every rule using codes defined in the rubric (e.g., ``O'' for open-ended and ``C'' for closed-ended questions, ``U'' for usage questions and ``I'' for improvement questions, and whether the question asks about a single item ``Y'' or two or more items ``N''). For the topic-ordering rule, the model is instructed to assign a rank to each topic based on how general the topic is in relation to other topics. Topics are flagged as unordered, if their rank and script index mismatch. For non-labeling rules, e.g., the script-length rule that requires more than one question per substantive topic and at least five substantive topics, the rule score is computed algorithmically. 

We ensure validity of script-quality judge's output by verifying its consistency with a human rater. The third author manually applied the same rubric to 1/3 of the dataset (15 scripts), after which we compute the mean Cohen's $\kappa$ = 0.68 across all rules, which indicates substantial agreement~\cite{Landis1977Measurement}. Thus, the judge scored the remaining 2/3 of the dataset.

The null hypothesis \textbf{H1} is that the total rubric score is drawn from the same distribution for scripts under AI-assisted and training-only conditions. We test the hypothesis using Welch's $t$-test, as recommended for SE experiments~\cite{arcuriHitchhikerGuide2014}; we report Hedges' $g$ and 95\% confidence intervals as measures of effect size. We analyze individual rubric dimensions as secondary outcomes and adjust p-values using the Benjamini–Hochberg procedure to control the false discovery rate.

We also report the goal alignment composition as a secondary analysis. Each question coded under the goal alignment rule targets either existing features \emph{usage} (U), a needed \emph{improvement} or missing feature (I), or \emph{neither} (N); the rubric credits a question as \textit{goal-aligned} when it is U or I. Because prior work does not prescribe an optimal ratio of usage and improvement questions in a script, we do not score the ratio. However, we compare the ratios between conditions and the degree to which usage and improvement questions are interleaved within the script, reporting descriptive effect sizes.

\subsection{RQ2: Measuring Interaction Quality}
\label{sec:rq2-metrics}

Research question RQ2 examines the difference between the AI-assisted and training-only transcripts using \textit{interaction volume}, which characterizes the overall amount of interviewer activity devoted to eliciting information, and \textit{interaction focus}, which characterizes how follow-up activity is distributed across topics and questions covered in the interview. 

We assign 3 binary codes (\emph{yes}/\emph{no}) to each interviewer turn:
\begin{itemize}
    \item \texttt{is\_scripted} --- whether the turn instantiates a question that appears in the interview script;
    \item \texttt{is\_follow\_up} --- whether the turn is a follow-up probing question to the stakeholder's preceding answer rather than a new scripted question;
    \item \texttt{is\_emergent} --- whether the turn is neither a scripted nor a follow-up question.
\end{itemize}

We tested two methods to label each transcript using \texttt{gpt-5.4-mini}: 30-turns-per-task, in which each dyad is segmented into 30-turn windows with a 4-turn overlap on both ends of the window; and 1-turn-per-task, in which the each dyad is segmented into a 1-turn window with 4-turn overlap on both ends. Because the per-turn method produced a higher error rate, we adopted the 30-turn method. We also prompted the LLM to return the index of the scripted question that was matched with an interviewer turn. 

We label interviewer turns using an \textit{interview-turn-label judge} based on LLM-as-a-Judge~\cite{zheng2023llmjudge} with \texttt{gpt-5.4-mini}, and validate the judge using a ground truth dataset created by the first author, who manually labeled each turn in 1/3 of the dataset (15 transcripts). The Cohen's $\kappa$ = 0.66, denoting substantial agreement between the human and judge~\cite{Landis1977Measurement}. Therefore, we used the interview-turn-label judge to label the remaining 1,066 interviewer turns in the dataset.

We test two hypotheses that capture complementary properties of interviewer behavior. 

\textbf{H2.1 Interaction Volume.} Interaction volume differs between AI-assisted and training-only interviews, measured by the number of interviewer turns, scripted questions covered, follow-up questions, emergent questions (questions about topics not part of the script), and topics covered.

Interaction volume is operationalized using five count-based indicators: (i) total interviewer turns, (ii) scripted questions covered, (iii) follow-up questions, (iv) emergent questions, and (v) number of topics covered. Differences between conditions are analyzed using negative-binomial regression models, incidence rate ratios (IRRs), 95\% confidence intervals, and p-values. 

\textbf{H2.2 Interaction Focus.} Interaction focus differs between AI-assisted and training-only interviews, measured by how follow-up questions are distributed across the other question types and the topics covered in the interview.

Interaction focus is operationalized using two primary ratios applied to each interview: (i) \emph{follow-ups per non-follow-up question}, which measures the ratio of follow-up questions to scripted and emergent questions covered in the interview-- these question types trigger the stakeholder responses that lead to follow-up questions -- computed by $F_q / (S_q + E_q)$ from the total follow-up $F_q$, scripted $S_q$ and emergent $E_q$ questions covered; and (ii) \emph{follow-ups per topic}, which measures elaboration within each discussed topic, computed by $F_q / T$ from the total follow-up questions $F_q$ and topics $T$. As secondary evidence of topic concentration, we also report \emph{interviewer turns per topic}. Because these ratios are reported on a per-interview basis, differences between conditions are analyzed using Welch's $t$-test, with Hedges' $g$ and 95\% confidence intervals reported as measures of effect size. 

\subsection{RQ3: Measuring Requirement Refinement}
\label{sec:rq3-metrics}

Research question RQ3 analyzes the interview outcome artifact, which is the elicited requirements expressed as goal models~\cite{anton1996gore, lamsweerde2001gore}. Goal models are robust requirements models because they capture high-level descriptions of environmental states, which are traceable directly to low-level system operations used to achieve and maintain those states. Thus, if an interview covers a mix of abstract stakeholder needs and concrete system functions to support those needs, a single goal model can represent all of these entities. We adopted the LLM-based \textit{goal model extraction} method proposed by Sharfuddin and Breaux~\cite{sharfuddinGenerativeGoalModeling2025}, which consists of a four-step pipeline~\footnote{Available online: https://github.com/cmu-relab/goalgeneration}: (1) extracting goals from a sliding transcript window by prompting an LLM, (2) validating extracted goals by tracing them back to transcript turns, (3) clustering semantically similar goals, and (4) constructing a goal-refinement graph. Their original work relied on \texttt{gpt-4o}, which we adapted to \texttt{gpt-5.4-mini}. The first author manually verified 15 LLM-generated goal models and found that goal refinements were 82\% accurate.

Goal models are directed, acyclic graphs in which each node represents a unique goal, directed edges represent refinement relationships from low-level goals to higher-level goals, and source nodes describe high-level, abstract requirements, while sink nodes describe low-level, concrete requirements. The total number of nodes represents breadth of the requirements elicited, and proportion of sink nodes to the total number of nodes represents depth. This is because sink nodes refine higher-level goals by answering ``how'' those high-level goals are achieved or maintained by the system.

We evaluate two complementary properties of the extracted goal models: \textit{requirement breadth}, which characterizes the amount of elicited requirements, and \textit{requirement depth}, which characterizes the extent to which those requirements are refined into concrete, low-level requirements.

\textbf{H3.1 Requirement Breadth.} Requirement breadth differs between AI-assisted and training-only interviews, measured by the number of goal nodes, refinement edges, and sink nodes in the extracted goal graph.

Requirement breadth is operationalized using three complementary graph-size measures: graph nodes, graph edges, and sink nodes. Differences between conditions are analyzed using negative-binomial regression models, reporting incidence rate ratios (IRRs), 95\% confidence intervals, and p-values.



\textbf{H3.2 Requirement Depth.} Requirement depth differs between AI-assisted and training-only interviews, measured by the proportion of sink nodes to total nodes in the goal graph.

Requirement depth is measured using lowest-level goal fraction (sink nodes / graph nodes), which is the proportion of requirements represented as low-level goals. Differences between conditions are analyzed using Welch's $t$-test, with Hedges' $g$ and 95\% confidence intervals for effect size.

\section{Results}
\label{sec:results}

We now present our results, organized by RQs.

\subsection{RQ1: Script Quality}

\begin{table}[t]
\centering
\scriptsize
\caption{Script-quality rubric rule scores (0--100\% compliance). $p$-values are Benjamini--Hochberg adjusted. AI refers to AI-assisted.}
\label{tab:script-critique}
\begin{tabular}{lrrr}
\hline
Rubric rule & AI $M$ & Train $M$ & $p$ \\
\hline
R1 Goal alignment & 97.2 & 84.7 & $<.001$ \\
R2 Introduction + rapport building & 100.0 & 90.9 & .482 \\
R3 Context-free opener & 100.0 & 68.2 & .013 \\
R4 Topic ordering & 78.4 & 90.9 & .001 \\
R5 Length ($>1$ Q/topic, $\geq 5$ topics) & 85.4 & 10.8 & $<.001$ \\
R6 Open-ended & 85.9 & 72.6 & $<.001$ \\
R7 One question per turn & 90.6 & 74.4 & .001 \\
R8 Clarity ($\le 30$ words) & 97.5 & 94.5 & .094 \\
R9 Concluding with a summary & 100.0 & 86.4 & .261 \\
\hline
Total (out of 100) & 92.8 & 74.8 & $<.001$ \\
\hline
\end{tabular}
\end{table}

Research question RQ1 asks whether the AI-generated interview scripts are the same, higher, or lower quality than human-authored scripts. Table~\ref{tab:script-critique} reports the per-rule compliance scores for the 21 AI-generated scripts and the 22 training-only scripts. AI-generated scripts score substantially higher overall: mean 92.8/100.0, compared with mean 74.8/100.0 for training-only scripts. The difference is large and statistically significant (Welch $t=9.84$, $p<.001$, Hedges' $g=2.89$, 95\% CI $[2.27, 3.97]$).

The individual rule-based scores provide diagnostic insight into the overall difference. AI-generated scripts predominantly include more than one question per topic and at least five topics (85.4) , whereas human-authored scripts frequently include only one question per topic or less than five topics (10.8). AI-generated scripts score higher on asking one question per turn (90.6 vs.\ 74.4), on goal alignment (97.2 vs.\ 84.7), on open-ended phrasing (85.9 vs.\ 72.6), and on opening with a context-free question (100.0 vs.\ 68.2); all of these differences persist under Benjamini--Hochberg correction. The exception is coherent topic ordering, where AI-generated scripts score lower than training-only scripts (78.4 vs.\ 90.9; $p=.001$ after correction).


As a secondary finding, the goal-alignment label distribution differs sharply between conditions. AI-assisted scripts apportion more questions to improvements or missing features (\%I $=44.1$ vs.\ $23.8$; Hedges' $g=2.13$, 95\% CI $[1.36, 3.30]$), and interleave usage and improvement questions more evenly (interleaving index $0.69$ vs.\ $0.19$; $g=3.29$, 95\% CI $[2.25, 5.26]$). Thus, the generated scripts not only satisfy formal rubric items, but they focus more narrowly on the interview goal, including on unmet needs and improvement opportunities.

\rqanswer{AI-generated scripts show substantially higher quality overall than training-only scripts under the rubric, especially for script length, open-endedness, and goal alignment. The main exception is topic ordering, wherein human-authored scripts are higher quality.}
\vspace{-3mm}
\subsection{RQ2: Interaction Quality}

\begin{table}[t]
\centering
\caption{Statistics for Scripts and Interactions. Effect sizes are AI relative to training-only, with 95\% CIs.}
\label{tab:turn-analysis}
\resizebox{\columnwidth}{!}{%
\begin{tabular}{lrrcr}
\hline
Metric & AI $M$ & Train $M$ & Effect [95\% CI] & $p$ \\
\hline
\multicolumn{5}{l}{\textit{Script design (IRR)}} \\
Scripted questions & 18.9 & 19.4 & 0.97 [0.83, 1.14] & .750 \\
Scripted topics & 8.7 & 17.2 & 0.51 [0.43, 0.60] & $.001$ \\
\hline
\multicolumn{5}{l}{\textit{H2.1 --- Interaction volume (IRR)}} \\
Interviewer turns & 54.5 & 32.1 & 1.70 [1.16, 2.49] & .006 \\
Scripted questions covered & 17.9 & 13.3 & 1.34 [1.14, 1.58] & .001 \\
Follow-up questions & 33.7 & 16.3 & 2.06 [1.15, 3.71] & .015 \\
Emergent questions & 2.4 & 2.2 & 1.09 [0.64, 1.86] & .750 \\
Topics covered & 9.6 & 14.5 & 0.66 [0.56, 0.77] & .001 \\
\hline
\multicolumn{5}{l}{\textit{H2.2 --- Interaction focus (Hedges' $g$)}} \\
Follow-ups / non-follow-up qn. & 1.61 & 1.07 & $+0.39$ [$-0.21$, $+0.80$] & .210 \\
Follow-ups / topic & 3.43 & 1.15 & $+0.81$ [$+0.61$, $+1.59$] & .015 \\
Interviewer turns / topic & 5.63 & 2.26 & $+1.18$ [$+0.94$, $+2.52$] & .001 \\
\hline
\end{tabular}%
}
\end{table}


\begin{figure}[t]
\centering
\includegraphics[width=\columnwidth]{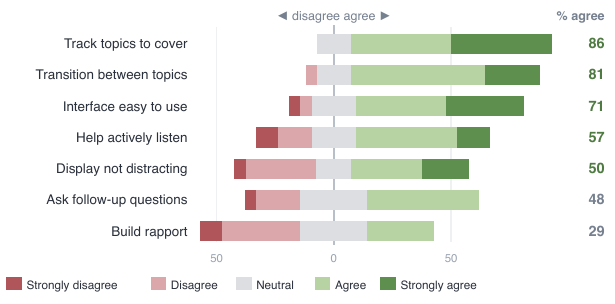}
\caption{AI-assisted interviewers' ratings of the assisted capabilities (5-point Likert). Bars show the full response distribution diverging from neutral; \% agree is the share selecting the two most favorable responses. }
\label{fig:tool-capabilities}
\vspace{-5mm}
\end{figure}

Research question RQ2 examines the differences in interaction volume and focus between AI-assisted and training-only interviews. Table~\ref{tab:turn-analysis} summarizes the results.

\textbf{H2.1 Interaction Volume.} Interaction volume differs between conditions. AI-assisted interviews average 54.5 interviewer turns, compared to 32.1 in training-only interviews (IRR $=1.70$, $p=.006$), with more scripted questions covered (17.9 vs.\ 13.3; IRR $=1.34$, $p=.001$) and more follow-up questions (33.7 vs.\ 16.3; IRR $=2.06$, $p=.015$). Emergent questions (neither scripted nor follow-up questions) are similar across conditions (2.4 vs.\ 2.2; $p=.75$), and AI-assisted interviews cover fewer distinct topics (9.6 vs.\ 14.5; IRR $=0.66$, $p=.001$). Thus, the AI-assisted condition produces more questioning activity distributed over fewer topics.


\textbf{H2.2 Interaction Focus.} Interaction focus differs between conditions. AI-assisted interviews show higher follow-ups per topic (3.43 vs.\ 1.15; Hedges' $g=+0.81$, $p=.015$) and more interviewer turns per topic (5.63 vs.\ 2.26; $g=+1.18$, $p=.001$). However, the ratio of follow-ups to covered non-follow-up questions (covered scripted and emergent) is not significantly different between conditions (1.61 vs.\ 1.07; $g=+0.39$, $p=.21$). Although the total number of follow-ups is significantly higher under AI-assisted (33.7 vs.\ 16.3), pointing to greater overall focus on probing during the interview, we cannot conclude that the AI-assisted condition distributes follow-ups evenly across questions: some questions seem to attract disproportionately more follow-up than others, and the turn-level data does not reveal what drives this imbalance.

Differences in interview script design correlate with interaction focus. Both conditions prepared a similar number of scripted questions (18.9 vs.\ 19.4; $p=.75$), but AI-generated scripts grouped them into fewer topic blocks (8.7 vs.\ 17.2; IRR $=0.51$, $p.001$). AI-assisted interviewers covered 9.6 distinct topics against 8.7 scripted topics; training-only interviewers covered 14.5 against 17.2 scripted topics. Expressed as the mean of the ratios of scripted questions covered per-interview, the AI-assisted condition averaged 86\% versus 69\% under training-only ($p=.002$).


\begin{figure}[t]
\centering
\includegraphics[width=\columnwidth,trim={46pt 8pt 28pt 60pt},clip]{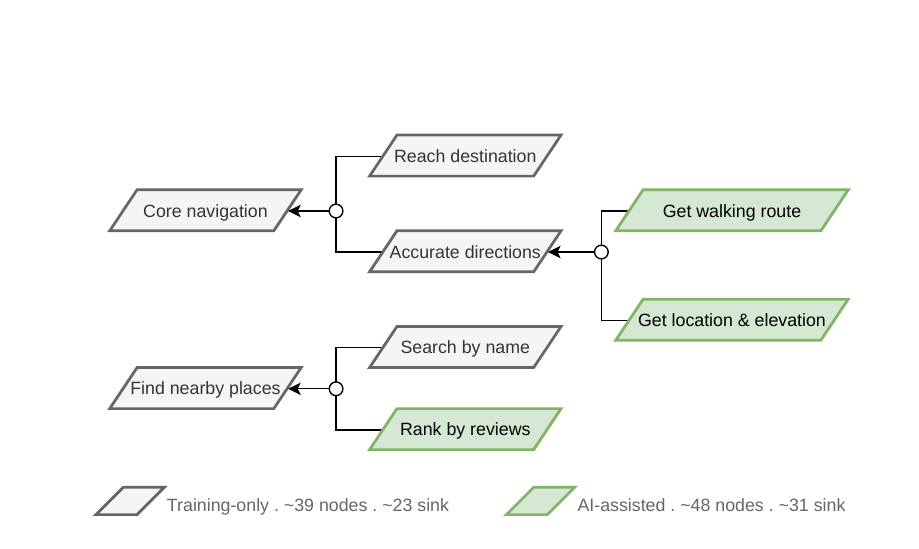}
\caption{Goal refinements from a navigation-app interview. Node labels are paraphrased; the node and sink counts are means over 43 interviews.}
\label{fig:goal-compare}
\vspace{-2mm}
\end{figure}

\textbf{System engagement (AI-assisted only).} We analyzed the post-interview survey responses from AI-assisted interviewers and the script management system's telemetry to understand how interviewers engage with the system. Figure~\ref{fig:tool-capabilities} summarizes the ratings of assisted capabilities on a 5-point Likert scale (1 = strongly disagree through 5 = strongly agree; 3 = neutral), with each item phrased as a declarative statement about whether a feature was useful; we report \% agreement as the share of responses with rating $\geq 4$. Topic-coverage tracking received the highest ratings ($M=4.29$; 86\% agreement), followed by support for transitioning between topics ($M=4.00$; 81\%). Ratings were lower for on-demand follow-up generation ($M=3.19$; 48\%) and lowest for rapport support ($M=2.76$; 29\%); interviewers were split on whether the interface distracted from the conversation (50\% agreement). Telemetry shows that 28/140 generated follow-ups ($\approx$20\%) were adopted verbatim by interviewers. The topic tracker agreed closely with post-hoc LLM coding of covered scripted topics (F1 $\approx 0.91$, precision $\approx 0.96$; see Section~\ref{sec:rq2-metrics} for coding details).


\rqanswer{The AI-assisted condition differs from training-only in interaction volume and per-topic focus. Follow-ups and interviewer turns per topic are significantly higher under AI assistance; the script-invariant follow-up rate (follow-ups per non-follow-up question) does not differ between conditions.}


\subsection{RQ3: Requirement Refinement}


Research question RQ3 analyzes the extracted goal graphs. Table~\ref{tab:goal-modeling-results} summarizes the results.

\textbf{H3.1 Requirement Breadth.} AI-assisted interviews produce larger goal graphs than training-only interviews (Table~\ref{tab:goal-modeling-results}). Goal graphs extracted from AI-assisted interviews contain more graph nodes (47.5 vs. 38.6), more refinement edges (68.8 vs. 51.0; $p=.003$), and more sink nodes (31.4 vs. 23.3; $p=.012$). The increase in graph nodes is only marginally significant ($p=.053$).

These findings should be interpreted with the interview-length difference in mind. AI-assisted interviews are substantially longer than training-only interviews (54.5 vs.\ 32.1 interviewer turns). AI-assisted interviews produce fewer goals per transcript turn than training-only interviews (0.98 vs. 1.27). Thus, the larger goal graphs reflect total interview activity rather than a higher rate of requirement extraction per turn.

\begin{table}[t]
\centering
\caption{RQ3: Goal-modeling results. Count metrics report negative-binomial IRRs; lowest-level goal fraction reports Hedges' $g$.}
\label{tab:goal-modeling-results}
\resizebox{\columnwidth}{!}{%
\begin{tabular}{lrrrr}
\hline
Metric & AI $M$ & Train $M$ & Effect [95\% CI] & $p$ \\
\hline
\multicolumn{5}{l}{\textit{H3.1 --- Breadth (IRR)}} \\
Graph nodes      & 47.5 & 38.6 & 1.23 [1.00, 1.52] & .053 \\
Graph edges      & 68.8 & 51.0 & 1.35 [1.11, 1.64] & .003 \\
Sink nodes       & 31.4 & 23.3 & 1.35 [1.07, 1.70] & .012 \\
\hline
\multicolumn{5}{l}{\textit{H3.2 --- Depth (Hedges' $g$)}} \\
Lowest-level goal fraction    & 0.653 & 0.598 & $+0.88$ [$+0.34$, $+1.54$] & .005 \\
\hline
\end{tabular}%
}
\vspace{-5mm}
\end{table}

\textbf{H3.2 Requirement Depth.} Goal models extracted from AI-assisted interviews contain a significantly higher lowest-level goal fraction than those extracted from training-only interviews (0.653 vs. 0.598; $p=.005$), with a large effect size. Although AI-assisted interviews show a non-significant increase in total number of graph nodes, they show a significantly higher proportion of sink nodes. This pattern suggests that the additional requirements elicited under the workflow are disproportionately represented as low-level requirements rather than leading to an equitable increase in both high- and low-level requirements. Figure~\ref{fig:goal-compare} illustrates this difference. While both conditions recover similar high-level stakeholder goals, goal models extracted from AI-assisted interviews refine these goals into a greater proportion of low-level goals.


\rqanswer{Goal models extracted from the AI-assisted transcripts are more refined than those from the training-only transcripts. AI-assisted interviews also yield larger goal graphs, but the node-count increase is only marginally significant and co-varies with interview length; we therefore treat breadth as inconclusive.}

\section{Discussion and Implications}
\label{sec:discussion}

We now discuss our results.

\subsection{From More Topic Coverage to Focused Elaboration}

The main study implication is that the AI-assisted and training-only workflows organize the interview around different assumptions about how to conduct the interview. AI-assisted interviewers begin with fewer topics, more questions per topic, and an interface that keeps those topics visible. Training-only interviewers begin with more topics and manage their own progression through the script. This initial difference shapes the interview: the AI-assisted workflow encourages interviewers to remain on-topic longer to refine those topics into more low-level requirements, while the training-only workflow supports the coverage of more topics with some coverage loss.

The results intersect the core issue of interview time management. A script with a large number of topics appears attractive because it promotes exploration and may increase the likelihood of discovering important requirements, whereas a smaller number of narrower topics is restrictive. Interviewing with many scripted topics can increase the interviewer's cognitive load: each topic competes for limited interview time, and the interviewer must decide whether to move on or to continue probing. The AI-assisted condition begins with a focused script that yields a higher proportion of low-level requirements. This suggests that a shorter topic list can be productive when the purpose is to obtain greater refinement in stakeholder needs. 

\textbf{Implications for practitioners and educators.} Practitioners should intentionally choose the number of scripted topics based on the interview duration and the desired depth of elicitation. Including too many topics may limit opportunities to probe stakeholder responses, whereas a smaller set of topics allows more time for follow-up questions and the elicitation of concrete requirements. Early exploratory interviews may therefore benefit from broader topic coverage, while later interviews can focus on a smaller set of high-priority topics in greater depth. For RE education, the results also distinguish script support from interviewing skill: AI assistance can help novices manage a focused interview script, but should not be assumed to teach when to depart from that script, pursue an unexpected line of inquiry, or improve topic ordering.

\textbf{Implications for tool builders.} Future AI assistance should help interviewers see the consequence of starting with a long or short topic list. A useful system would help the interviewer decide whether the interview currently needs more topic coverage or deeper elaboration on fewer topics. 

\subsection{Decomposing the Workflow Outcomes}

A key question is whether the observed differences reflect offline script generation, online script management, or both. Although the quasi-experiment does not isolate these phases, the RQ2 results constrain how we interpret interaction focus and system use.

First, per-topic focus metrics could partly reflect script design. On average, AI-generated scripts exhibited more questions in fewer topic blocks ($\approx$2.2 vs.\ 1.1 questions per block), a higher proportion of improvement questions per script (44.1 vs.\ 23.8), and AI-assisted interviewers covered more of their scripts (86\% vs.\ 69\%). While AI-assisted interviews covered fewer topics (9.6 vs.\ 14.5), they produced more follow-up questions (33.7 vs.\ 16.3). Dividing that larger follow-up total by a smaller topic count leads to more follow-ups and interviewer turns \emph{per topic} (3.43 vs.\ 1.15; 5.63 vs.\ 2.26). Because AI-generated scripts cover fewer topics than human-authored scripts, the difference in topics could naturally inflate the values of the per-topic metrics in favor of the AI-assisted condition. However, the increase in the number of follow-up questions cannot be explained only by the larger topic blocks---it could be a consequence of the larger proportion of improvement questions, the topic tracking, or follow-up question prompting. 


Second, survey and telemetry data characterize how interviewers used online script management. Topic tracking was rated most useful ($M=4.29$/5; 86\% agreement), while follow-up recommendations were rarely adopted verbatim ($\approx$20\%). Interviewers reported that displayed follow-ups were often stale---interface latency meant suggestions referred to topics already covered---prompting them to formulate their own follow-ups instead. The generator therefore appears to function less as a source of verbatim questions and more as a cognitive prompt that encourages interviewers to formulate their own follow-up questions, which helps reconcile low adoption with the observed increase in follow-up volume.

With these patterns in mind, evidence points more strongly to integrating offline script generation with topic tracking than with follow-up generation. The value of the workflow lies in making a selected topic space easier to maintain during the interview, while generated follow-ups only play a supporting role in this study. As interviewers noted, a key limitation of follow-up generation was stale suggestions and display latency; improvements in speed, relevance, and interviewer-style phrasing may increase this capability's utility.


\textbf{Implications for researchers.} This study establishes a practical baseline for AI-assisted requirements elicitation using off-the-shelf language models without task-specific fine-tuning. Future research should investigate how to generate more relevant follow-up questions within the time constraints of live interviews, for example through smaller, faster models or methods that better identify opportunities where follow-up questions lead to meaningful design insight.


\section{Threats to Validity}
\label{sec:threats-validity}

\textit{Construct validity} addresses whether what we measure is actually the construct of interest~\cite{Yin2017CaseStudy}. The outcomes in this study are indirect measures of elicitation quality rather than direct evidence that the resulting requirements are correct, complete, or useful to engineers. The rubric used to measure script quality lists desirable script properties derived from prior elicitation literature, but not whether every question would elicit new requirements from a real stakeholder. Furthermore, the LLM-generated goal models used to answer RQ3 could contain inaccurate goal refinements. We incorporated the goal-model generation technique from prior work~\cite{sharfuddinGenerativeGoalModeling2025}, who reported a 72.2\% accuracy on identifying goal refinements using \texttt{GPT-4o}. In this study, the first author manually verified 15 \texttt{gpt-5.4-mini}-generated goal models and found that the refinements were 82\% accurate. 

The study relies on LLM-as-a-Judge to evaluate the RQs. The script-quality judge, interviewer-turn-label judge, and goal model extraction procedures may introduce systematic errors or encode model-specific biases. We mitigate this threat by validating the script judge and turn-level coder using expert labels and reporting above-chance agreement using Cohen's $\kappa$. In addition, participants in the training-only group could have used AI to author scripts, despite prohibitive instructions in the study, which could reduce the script differences.

\textit{Internal validity} concerns whether effects follow from the causes in an empirical study~\cite{Yin2017CaseStudy}. Unlike a controlled experiment with random assignment of participants drawn from the same population where an intervention is the single variable, we employ a quasi-experimental design to compare two settings: interviewers who received no training but used the AI-assisted workflow; and interviewers who received course-based RE training but did not use the workflow. Consequently, observed differences may reflect the workflow, training, cohort differences, selection effects, or prior interviewing experience. The study cannot attribute gains independently to offline script generation or online script management.

The two settings also differ in the structure of the interview scripts. While AI-generated scripts have more questions per topic, the training-only scripts include more topics and fewer questions on each topic. To isolate the effect of this difference on interaction focus, we report a script-invariant rate (follow-ups per non-follow-up question asked); this shows no difference between conditions. Therefore, the evidence for the AI-assisted condition leading to a difference in the probing intensity per question remains inconclusive. 

The study sample is small, with approximately 21 interviewers per condition, and is powered primarily to detect large effect sizes. Statistically insignificant results are therefore inconclusive rather than evidence of no effect. We report effect sizes and confidence intervals alongside p-values. Findings describe evidence of the direction of the workflow's effects, not as precise estimates of population-level magnitudes.

\textit{External validity} refers to the extent to which we can generalize the results~\cite{Yin2017CaseStudy}. The study uses graduate students as interviewers and stakeholders. While all participants had at least one year of prior industrial software engineering experience, this experience would still amount to an entry-level software developer. In addition, the business goals were simulated and could be new to participants in the AI-assisted condition, which would be different in an industrial setting with regular internal messaging on business objectives. The simulated goals were acquired from public SEC 10-K disclosures, which may not represent private, non-profit, and government organizations, and which may not cover the full breadth and depth of internal business goals. The app categories are all web and mobile apps, whereas other application types, such as cyber-physical or embedded systems, may not produce the same effects in either condition. 



The workflow and evaluation use OpenAI's frontier models \texttt{gpt-5-mini} and \texttt{gpt-5.4-mini}, which are both large reasoning models trained to generate long-chain of thought. We did not experiment with other reasoning models (e.g., Claude Sonnet 3.7 or later), non-reasoning models (e.g., GPT-4.1) or open source models. Frontier models will change over time, performance can vary widely due to their unique training procedures, and repeated runs may differ due to model updates. This affects the reproducibility of script generation, transcript correction, LLM judging, and goal-model extraction.

Finally, participants may have behaved differently because they knew they were using AI assistance or because they were being observed in a research study. Such demand and novelty effects can increase engagement, encourage closer adherence to the interface, or change how interviewers use their scripts~\cite{cambridge2012effects}. We accept this threat as a limitation of the study design. In practice, results may differ as the novelty of AI assistance diminishes, as interviewers become more accountable for downstream requirements decisions, or as practitioners bring stronger positive or negative expectations about AI into the interview.

\section{Conclusion}
\label{sec:conclusion}
We present an AI-assisted elicitation workflow and evaluate it against a training-only condition in a quasi-experimental study. Across script quality, interaction quality, and extracted goal-model refinements, the results show that AI-generated scripts are more aligned with interviewing guidance, AI-assisted interviews include more follow-up questions and topic-focused interaction, and the resulting goal models contain a higher proportion of low-level requirements.

However, the findings do not support a ``more AI is better'' conclusion. The AI-assisted condition exhibited a different interview trajectory: more turns over fewer topics, guided by a focused script. Within that condition, interviewers achieved greater script coverage, asked more follow-up questions, and the resulting goal models were more deeply refined. Interviewers rated topic tracking as the most useful online script management capability (mean 4.29/5; 86\% agreement), whose separate behavioral effect remains to be isolated. These findings motivate workflows of this kind as candidate elicitation platforms for balancing coverage, depth, and adaptation, pending designs that can isolate their components.

\bibliographystyle{IEEEtran}
\bibliography{references}

@article{sommerville1999benefitsRE,
  author={Sawyer, P. and Sommerville, I. and Viller, S.},
  journal={IEEE Software}, 
  title={Capturing the benefits of requirements engineering}, 
  year={1999},
  volume={16},
  number={2},
  pages={78-85},
  doi={10.1109/52.754057}}

@inproceedings{lamsweerde2001gore,
author = {Van Lamsweerde, Axel},
title = {Goal-Oriented Requirements Engineering: A Guided Tour},
year = {2001},
publisher = {IEEE Computer Society},
address = {USA},
booktitle = {Proceedings of the Fifth IEEE International Symposium on Requirements Engineering},
pages = {249},
series = {RE '01}
}

@INPROCEEDINGS{anton1996gore,
  author={Anton, A.I.},
  booktitle={Proceedings of the Second International Conference on Requirements Engineering}, 
  title={Goal-based requirements analysis}, 
  year={1996},
  volume={},
  number={},
  pages={136-144},
  doi={10.1109/ICRE.1996.491438}}

@article{arcuriHitchhikerGuide2014,
  title     = {A Hitchhiker's Guide to Statistical Tests for Assessing Randomized Algorithms in Software Engineering},
  author    = {Arcuri, Andrea and Briand, Lionel},
  journal   = {Software Testing, Verification and Reliability},
  volume    = {24},
  number    = {3},
  pages     = {219--250},
  year      = {2014},
  publisher = {Wiley}
}

@incollection{zowghiRequirementsElicitationSurvey2005,
  title     = {Requirements Elicitation: A Survey of Techniques, Approaches, and Tools},
  author    = {Zowghi, Didar and Coulin, Chad},
  booktitle = {Engineering and Managing Software Requirements},
  editor    = {Aurum, A. and Wohlin, C.},
  publisher = {Springer},
  address   = {Berlin, Heidelberg},
  pages     = {19--46},
  year      = {2005}
}

@inproceedings{davisEffectivenessRequirementsElicitation2006,
  title     = {Effectiveness of Requirements Elicitation Techniques: Empirical Results Derived from a Systematic Review},
  author    = {Davis, Alan and Dieste, Oscar and Hickey, Ann and Juristo, Natalia and Moreno, Ana M.},
  booktitle = {14th IEEE International Requirements Engineering Conference (RE'06)},
  pages     = {179--188},
  year      = {2006},
  doi       = {10.1109/RE.2006.17}
}

@article{palomaresStateofpracticeRequirements2021,
  title   = {The State-of-Practice in Requirements Elicitation: An Extended Interview Study at 12 Companies},
  author  = {Palomares, Cristina and Franch, Xavier and Quer, Carme and Chatzipetrou, Panagiota and L{\'o}pez, Lidia and Gorschek, Tony},
  journal = {Requirements Engineering},
  volume  = {26},
  pages   = {273--299},
  year    = {2021}
}

@article{ferrariAmbiguityTacitKnowledge2016,
  title   = {Ambiguity and Tacit Knowledge in Requirements Elicitation Interviews},
  author  = {Ferrari, Alessio and Spoletini, Paola and Gnesi, Stefania},
  journal = {Requirements Engineering},
  volume  = {21},
  number  = {3},
  pages   = {333--355},
  year    = {2016}
}

@article{hadarRoleDomainKnowledge2014,
  title   = {The Role of Domain Knowledge in Requirements Elicitation via Interviews: An Exploratory Study},
  author  = {Hadar, Irit and Soffer, Pnina and Kenzi, Keren},
  journal = {Requirements Engineering},
  volume  = {19},
  number  = {2},
  pages   = {143--159},
  year    = {2014}
}

@inproceedings{banoLearningMistakesEmpirical2018,
  title     = {Learning from Mistakes: An Empirical Study of Elicitation Interviews Performed by Novices},
  author    = {Bano, Muneera and Zowghi, Didar and Ferrari, Alessio and Spoletini, Paola and Donati, Beatrice},
  booktitle = {2018 IEEE 26th International Requirements Engineering Conference (RE)},
  pages     = {182--193},
  year      = {2018}
}

@inproceedings{donatiCommonMistakesStudent2017,
  title     = {Common Mistakes of Student Analysts in Requirements Elicitation Interviews},
  author    = {Donati, Beatrice and Ferrari, Alessio and Spoletini, Paola and Gnesi, Stefania},
  booktitle = {Requirements Engineering: Foundation for Software Quality (REFSQ)},
  volume    = {10153},
  pages     = {148--164},
  year      = {2017}
}

@inproceedings{spoletiniInterviewReviewEmpirical2018,
  title     = {Interview Review: An Empirical Study on Detecting Ambiguities in Requirements Elicitation Interviews},
  author    = {Spoletini, Paola and Ferrari, Alessio and Bano, Muneera and Zowghi, Didar and Gnesi, Stefania},
  booktitle = {Requirements Engineering: Foundation for Software Quality (REFSQ)},
  pages     = {101--118},
  year      = {2018}
}

@inproceedings{zalewskiCognitiveBiases2020,
  title     = {On Cognitive Biases in Requirements Elicitation},
  author    = {Zalewski, Andrzej and Borowa, Klara and Kowalski, Damian},
  booktitle = {Integrating Research and Practice in Software Engineering},
  publisher = {Springer},
  pages     = {111--123},
  year      = {2020}
}

@article{hidellaarachchiEffectsHumanAspects2022,
  title   = {The Effects of Human Aspects on the Requirements Engineering Process: A Systematic Literature Review},
  author  = {Hidellaarachchi, Dulaji and Grundy, John and Hoda, Rashina and Madampe, Kashumi},
  journal = {IEEE Transactions on Software Engineering},
  volume  = {48},
  number  = {6},
  pages   = {2105--2127},
  year    = {2022}
}

@phdthesis{hanwayEffectsCognitiveLoad2020,
  title  = {The Effects of Cognitive Load for Investigative Interviewers},
  author = {Hanway, Pamela},
  school = {University of Portsmouth},
  year   = {2020}
}

@article{lendingRubricEvaluateEnhance2022,
  title   = {A Rubric to Evaluate and Enhance Requirements Elicitation Interviewing Skills},
  author  = {Lending, Diane and Ezell, Jeremy D. and Dillon, Thomas W. and May, Jeffrey},
  journal = {Journal of Information Systems Education},
  volume  = {33},
  number  = {4},
  pages   = {371--387},
  year    = {2022}
}

@article{mohedasRecommendedInterviewingPractices2022,
  title   = {The Use of Recommended Interviewing Practices by Novice Engineering Designers to Elicit Information During Requirements Development},
  author  = {Mohedas, Ibrahim and Daly, Shanna R. and Loweth, Robert P. and Huynh, Linh and Cravens, Grace L. and Sienko, Kathleen H.},
  journal = {Design Science},
  volume  = {8},
  pages   = {e16},
  year    = {2022}
}

@inproceedings{zarembaTypologyQuestions2021,
  title     = {Towards a Typology of Questions for Requirements Elicitation Interviews},
  author    = {Zaremba, Olga and Liaskos, Sotirios},
  booktitle = {2021 IEEE 29th International Requirements Engineering Conference (RE)},
  pages     = {384--389},
  year      = {2021},
  doi       = {10.1109/RE51729.2021.00042}
}

@article{pittsImprovingRequirementsElicitation2007,
  title   = {Improving Requirements Elicitation: An Empirical Investigation of Procedural Prompts},
  author  = {Pitts, Mitzi G. and Browne, Glenn J.},
  journal = {Information Systems Journal},
  volume  = {17},
  pages   = {89--110},
  year    = {2007}
}

@article{wahbehSociotechnicalProcessQuestionnaire2020,
  title   = {A Socio-Technical-Based Process for Questionnaire Development in Requirements Elicitation via Interviews},
  author  = {Wahbeh, Abdullah and Sarnikar, Surendra and El-Gayar, Omar},
  journal = {Requirements Engineering},
  volume  = {25},
  number  = {3},
  pages   = {295--315},
  year    = {2020},
  doi     = {10.1007/s00766-019-00324-x}
}

@article{burnayWhatStakeholdersWill2014,
  title   = {What Stakeholders Will or Will Not Say: A Theoretical and Empirical Study of Topic Importance in Requirements Engineering Elicitation Interviews},
  author  = {Burnay, Corentin and Jureta, Ivan J. and Faulkner, St{\'e}phane},
  journal = {Information Systems},
  volume  = {46},
  pages   = {61--81},
  year    = {2014}
}

@inproceedings{ferrariLearningRequirementsElicitation2019,
  title     = {Learning Requirements Elicitation Interviews with Role-Playing, Self-Assessment and Peer-Review},
  author    = {Ferrari, Alessio and Spoletini, Paola and Bano, Muneera and Zowghi, Didar},
  booktitle = {2019 IEEE 27th International Requirements Engineering Conference (RE)},
  year      = {2019}
}

@article{ferrariSaPeerReverseSaPeer2020,
  title   = {{SaPeer} and {ReverseSaPeer}: Teaching Requirements Elicitation Interviews with Role-Playing and Role Reversal},
  author  = {Ferrari, Alessio and Spoletini, Paola and Bano, Muneera and Zowghi, Didar},
  journal = {Requirements Engineering},
  volume  = {25},
  number  = {4},
  pages   = {417--438},
  year    = {2020}
}

@inproceedings{zowghiTeachingRequirementsEngineering2003,
  title     = {Teaching Requirements Engineering Through Role Playing: Lessons Learnt},
  author    = {Zowghi, Didar and Paryani, S.},
  booktitle = {Proceedings. 11th IEEE International Requirements Engineering Conference},
  pages     = {233--241},
  year      = {2003}
}

@inproceedings{daunSurveyInstructionalApproaches2021,
  title     = {A Survey of Instructional Approaches in the Requirements Engineering Education Literature},
  author    = {Daun, Marian and Grubb, Alicia M. and Tenbergen, Bastian},
  booktitle = {2021 IEEE 29th International Requirements Engineering Conference (RE)},
  pages     = {257--268},
  year      = {2021}
}

@inproceedings{liuBARADynamicStatebased2023,
  title     = {{BARA}: A Dynamic State-Based Serious Game for Teaching Requirements Elicitation},
  author    = {Liu, Yang and Li, Tong and Huang, Zhen and Yang, Zhi},
  booktitle = {2023 IEEE/ACM 45th International Conference on Software Engineering: Software Engineering Education and Training (ICSE-SEET)},
  pages     = {141--152},
  year      = {2023}
}

@inproceedings{ibrahimDesignDevelopmentSerious2019,
  title     = {Design and Development of a Serious Game for the Teaching of Requirements Elicitation and Analysis},
  author    = {Ibrahim, Z. and Soo, M. C. and Soo, M. T. and Aris, H.},
  booktitle = {2019 IEEE International Conference on Engineering, Technology and Education (TALE)},
  pages     = {1--8},
  year      = {2019}
}

@inproceedings{gorerGeneratingRequirementsElicitation2023,
  title     = {Generating Requirements Elicitation Interview Scripts with Large Language Models},
  author    = {G{\"o}rer, Binnur and Aydemir, Fatma Ba{\c s}ak},
  booktitle = {2023 IEEE 31st International Requirements Engineering Conference Workshops (REW)},
  pages     = {44--51},
  year      = {2023},
  doi       = {10.1109/REW57809.2023.00015}
}

@inproceedings{gorerGPTPoweredElicitationInterview2024,
  title     = {{GPT}-Powered Elicitation Interview Script Generator for Requirements Engineering Training},
  author    = {G{\"o}rer, Binnur and Aydemir, Fatma Ba{\c s}ak},
  booktitle = {2024 IEEE 32nd International Requirements Engineering Conference (RE)},
  pages     = {372--379},
  year      = {2024}
}

@article{parkerGuidelinesIntegrationLarge2023,
  title   = {Guidelines for the Integration of Large Language Models in Developing and Refining Interview Protocols},
  author  = {Parker, Jessica L. and Richard, Veronica M. and Becker, Kimberly},
  journal = {The Qualitative Report},
  year    = {2023}
}

@inproceedings{ronankiInvestigatingChatGPTPotential2023,
  title     = {Investigating {ChatGPT}'s Potential to Assist in Requirements Elicitation Processes},
  author    = {Ronanki, Krishna and Berger, Christian and Horkoff, Jennifer},
  booktitle = {2023 49th Euromicro Conference on Software Engineering and Advanced Applications (SEAA)},
  pages     = {354--361},
  year      = {2023}
}

@inproceedings{shenRequirementsElicitationFollowup2025,
  title     = {Requirements Elicitation Follow-Up Question Generation},
  author    = {Shen, Yuchen and Singhal, Anmol and Breaux, Travis},
  booktitle = {2025 IEEE 33rd International Requirements Engineering Conference (RE)},
  year      = {2025}
}

@misc{zhangHarnessingPowerAI2025,
  title         = {Harnessing the Power of {AI} in Qualitative Research: Role Assignment, Engagement, and User Perceptions of {AI}-Generated Follow-Up Questions in Semi-Structured Interviews},
  author        = {Zhang, He and Liu, Yueyan and Guan, Xin and Cai, Jie and Carroll, John M.},
  year          = {2025},
  eprint        = {2509.13234},
  archivePrefix = {arXiv}
}

@inproceedings{liuInterviewAIssistantDesigning2025,
  title     = {Interview {AI}-ssistant: Designing for Real-Time Human-{AI} Collaboration in Interview Preparation and Execution},
  author    = {Liu, Zhe},
  booktitle = {Companion Proceedings of the 30th International Conference on Intelligent User Interfaces (IUI '25 Companion)},
  year      = {2025}
}

@misc{kornLLMREIAutomatingRequirements2025,
  title         = {{LLMREI}: Automating Requirements Elicitation Interviews with {LLMs}},
  author        = {Korn, Alexander and Gorsch, Samuel and Vogelsang, Andreas},
  year          = {2025},
  eprint        = {2507.02564},
  archivePrefix = {arXiv}
}

@article{bajajMUCEMultilingualUsecase2022,
  title   = {{MUCE}: A Multilingual Use Case Model Extractor Using {GPT}-3},
  author  = {Bajaj, Deepali and Goel, Anita and Gupta, S. C. and Batra, Hunar},
  journal = {International Journal of Information Technology},
  volume  = {14},
  number  = {3},
  pages   = {1543--1554},
  year    = {2022}
}

@inproceedings{chenUseGPT4Creating2023,
  title     = {On the Use of {GPT-4} for Creating Goal Models: An Exploratory Study},
  author    = {Chen, Boqi and Chen, Kua and Hassani, Sajjad and Yang, Yi and others},
  booktitle = {2023 IEEE 31st International Requirements Engineering Conference Workshops (REW)},
  pages     = {262--271},
  year      = {2023}
}

@inproceedings{sharfuddinGenerativeGoalModeling2025,
  title     = {Generative Goal Modeling},
  author    = {Sharfuddin, Aaron and Breaux, Travis D.},
  booktitle = {2025 IEEE 33rd International Conference on Requirements Engineering},
  year      = {2025}
}

@inproceedings{ElMaarawiTefur2026Debugging,
  author    = {Aref El-Maarawi Tefur and
               Farnaz Fotrousi and
               Alessio Ferrari and
               Paola Spoletini and
               Walid Maalej},
  title     = {Debugging Requirements Interview Scripts: A Framework for Quality Assessment},
  booktitle = {Proceedings of the 34th IEEE International Requirements Engineering Conference (RE 2026)},
  year      = {2026},
  publisher = {IEEE},
  address   = {},
  pages      = {},
  doi        = {},
}

@article{lavazza2013businessgoals,
author = {Lavazza, Luigi},
title = {Business goals, user needs, and requirements: A problem frame-based view},
journal = {Expert Systems},
volume = {30},
number = {3},
pages = {215-232},
doi = {https://doi.org/10.1111/j.1468-0394.2012.00648.x},
url = {https://onlinelibrary.wiley.com/doi/abs/10.1111/j.1468-0394.2012.00648.x},
eprint = {https://onlinelibrary.wiley.com/doi/pdf/10.1111/j.1468-0394.2012.00648.x},
year = {2013}
}

@book{Gause1989Exploring,
  author    = {Donald C. Gause and Gerald M. Weinberg},
  title     = {Exploring Requirements: Quality Before Design},
  publisher = {Dorset House Publishing},
  address   = {New York, NY},
  year      = {1989},
  isbn      = {9780932633132},
}

@ARTICLE{potts1994inquiry,
  author={Potts, C. and Takahashi, K. and Anton, A.I.},
  journal={IEEE Software}, 
  title={Inquiry-based requirements analysis}, 
  year={1994},
  volume={11},
  number={2},
  pages={21-32},
  doi={10.1109/52.268952}}

@inproceedings{zheng2023llmjudge,
author = {Zheng, Lianmin and Chiang, Wei-Lin and Sheng, Ying and Zhuang, Siyuan and Wu, Zhanghao and Zhuang, Yonghao and Lin, Zi and Li, Zhuohan and Li, Dacheng and Xing, Eric P. and Zhang, Hao and Gonzalez, Joseph E. and Stoica, Ion},
title = {Judging LLM-as-a-judge with MT-bench and Chatbot Arena},
year = {2023},
publisher = {Curran Associates Inc.},
address = {Red Hook, NY, USA},
booktitle = {Proceedings of the 37th International Conference on Neural Information Processing Systems},
articleno = {2020},
numpages = {29},
location = {New Orleans, LA, USA},
series = {NIPS '23}
}

@article{Landis1977Measurement,
  author  = {Landis, J. Richard and Koch, Gary G.},
  title   = {The Measurement of Observer Agreement for Categorical Data},
  journal = {Biometrics},
  volume  = {33},
  number  = {1},
  pages   = {159--174},
  year    = {1977},
  doi     = {10.2307/2529310},
}

@book{Yin2017CaseStudy,
  author    = {Robert K. Yin},
  title     = {Case Study Research and Applications: Design and Methods},
  edition   = {6},
  publisher = {SAGE Publications},
  address   = {Thousand Oaks, CA},
  year      = {2017},
  isbn      = {9781506336169},
}

@article{cambridge2012effects,
  title = {The Effects of Demand Characteristics on Research Participant Behaviours in Non-Laboratory Settings: A Systematic Review},
  author = {McCambridge, Jim and {de Bruin}, Marijn and Witton, John},
  year = 2012,
  month = jun,
  journal = {PLOS ONE},
  volume = {7},
  number = {6},
  pages = {1--6},
  publisher = {Public Library of Science},
  doi = {10.1371/journal.pone.0039116}
}

@article{storey2025software,
author = {Abrah\~{a}o, Silvia and Grundy, John and Pezz\`{e}, Mauro and Storey, Margaret-Anne and Tamburri, Damian A.},
title = {Software Engineering by and for Humans in an AI Era},
year = {2025},
issue_date = {June 2025},
publisher = {Association for Computing Machinery},
address = {New York, NY, USA},
volume = {34},
number = {5},
issn = {1049-331X},
url = {https://doi.org/10.1145/3715111},
doi = {10.1145/3715111},
journal = {ACM Trans. Softw. Eng. Methodol.},
month = may,
articleno = {129},
numpages = {46}
}

@article{zachariasDevelopersDilemmaOpportunities2026,
  title = {Developers' {{Dilemma}}: {{Opportunities}} and {{Pitfalls}} of {{Generative AI}} for {{Software Development}}},
  author = {Zacharias, Jan and Popova, Alina and von Zahn, Moritz and Chen, Johannes and Hinz, Oliver},
  year = 2026,
  month = mar,
  journal = {Business \& Information Systems Engineering},
  issn = {1867-0202},
  doi = {10.1007/s12599-026-00998-y}
}

@article{Jonsson2007Rubrics,
  author  = {Anders Jonsson and Gunilla Svingby},
  title   = {The Use of Scoring Rubrics: Reliability, Validity and Educational Consequences},
  journal = {Educational Research Review},
  volume  = {2},
  number  = {2},
  pages   = {130--144},
  year    = {2007},
  issn    = {1747-938X},
  doi     = {10.1016/j.edurev.2007.05.002},
}
\vspace{12pt}
\color{red}

\end{document}